\documentclass[a4paper, amsfonts, amssymb, amsmath]{article}
\usepackage[utf8]{inputenc}
\usepackage{color}
\usepackage[dvipsnames]{xcolor}
\usepackage{xfrac}

\usepackage{appendix}
\usepackage{soul}
\usepackage{ulem}
\usepackage{graphicx}
\usepackage{comment}
\usepackage{soul}
\usepackage{authblk}
\usepackage{amsmath,amssymb,amsfonts,graphicx}
\usepackage[a4paper,top=2cm,bottom=2cm,left=2cm,right=2cm]{geometry}%
\usepackage{tikz}

\title{The Phase Transition in the holographic Hard-Wall  model}

\author[1]{Matteo Rinaldi 
 \footnote{Corresponding author email: matteo.rinaldi@pg.infn.it} }
    \affil{INFN section of Perugia. Via A. Pascoli, Perugia, 06123, Italy.}

\author{Vicente Vento}

\affil{Departamento de F\'{\i}sica Te\'orica-IFIC, Universidad de Valencia- CSIC,
46100 Burjassot (Valencia), Spain.}

\begin{document}
\maketitle

\begin{abstract}

A Hawking-Page phase transition between AdS thermal and AdS Black Hole 
was presented as a mechanism for explaining the QCD deconfinement phase transition within 
holographic models.  In order to implement  temperature dependence in the confined phase we use a hard-wall AdS/QCD model, where the geometry at low temperatures is described also by a Black Hole metric. We then investigate the temperature dependence of glueball states described as gravitons propagating in deformed background spaces.  {Finally, we use  potential models to physically describe  the implications of our study}.

\end{abstract}


\maketitle

\section{Introduction}

A successful strategy for applying the  AdS/CFT correspondence and holography
\cite{Maldacena:1997re,Witten:1998zw}
to hadron physics is the so-called
bottom-up  approach. In this framework, one  starts from some non perturbative
features of QCD and
attempts to construct its five-dimensional holographic dual.  The
duality is implemented in nearly conformal conditions where QCD is defined on the four dimensional 
boundary.
Moreover, the confinement feature of QCD can be realized by
  introducing a bulk space which is a slice of $AdS_5$ whose size is 
related to $z_0 \sim 1/\Lambda_{QCD}$
\cite{Polchinski:2000uf,Brodsky:2003px,Erlich:2005qh,DaRold:2005mxj,
BoschiFilho:2005yh}. This is
the so called hard-wall (HW) approximation. 
We have recently proposed the calculations of the spectrum of the scalar and
tensor glueballs under the
assumption that, in this holographic approach,
the dual operator to the glueballs
could be  the
graviton, the latter thus plays a significant role to describe the lowest
lying
glueballs. The main result of our investigations is that
 we do not need to introduce
additional fields into any $AdS_5$ to describe the glueballs, the gravitons indeed
satisfy the duality boundary conditions and are able to describe the elementary
scalar and tensor glueball spectra  \cite{Rinaldi:2017wdn,Rinaldi:2021dxh}. 

Given the extensive experimental search and theoretical description of the 
deconfinement phase transition \cite{Aoki:2006we,Guenther:2022hmv}, holographic models could not 
remain silent. A traditional description has been carried out via a Hawking-Phase phase transition from an AdS thermal phase at low temperatures to a  Black Hole (BH) phase at high
temperatures~\cite{Herzog:2006ra}. Much research has been carried out to determine the deconfinement temperature and the behaviour of the glueball and meson
spectra after the phase transition {~\cite{Herzog:2006ra,Kajantie:2006hv,Colangelo:2009ra,Braga:2017apr,Gursoy:2009jd,Miranda:2009uw}}. Recently we have studied also the
 deconfinement phase transition in a HW  AdS/QCD model from the perspective of our description of the scalar and tensor glueball spectrum by analyzing the graviton in a BH metric \cite{Rinaldi:2021xat}. We found that the deconfinement phase in the Herzog approach \cite{Herzog:2006ra} is reached via a two steps mechanism, a conclusion    shared with other studies \cite{Shuryak:2003ty}  and holographic analyses \cite{Mateos:2006nu}. {However, the phase transition  in these studies was  first order or second order while the experimental data seem to point out that at high temperatures QCD deconfines  via a crossover mechanism~\cite{Guenther:2022hmv}.  }
 
Here we propose a different scenario based on the behavior of the scalar and tensor glueballs {at $T \neq 0$},
in $AdS_5$, but only assuming a unique BH metric, for the background, for both the low and high temperature regimes.
  We consider two scales: $z_0$ the confinement one, and $z_h$ the BH horizon.  For  $z_h>z_0$ we impose  Dirichlet or Neumann conditions on the mode functions at $z_0$, as in the usual HW model. This region resembles the $AdS$ thermal phase at low temperatures, and differs from it when $z_h$ is close to $z_0$. For any $z_h$  we have also a BH solution for the glueball modes and this solutions define the high temperature deconfining behavior. Since we implement boundary conditions in the confined metric a free energy differences appears between the two phases which generates a first order phase transition. { Since the $AdS$ model does not predict} deconfined states we implement its properties in  potential models to discuss deconfinement. In these models the dynamics of the temperature dependence is determined by the AdS-BH dynamics.

\section{An $AdS_5$ model for confinement}

We would like to study behavior of QCD as the temperature increases by studying its $AdS_5$ dual. For doing so we will study the behavior of the glueball spectrum with temperature. At $T=0$ the traditional  HW model reproduces quite well this spectrum \cite{Rinaldi:2017wdn}. In previous calculations two different metrics where used an $AdS$ thermal metric in the low $T$ phase and an $AdS$ black hole (BH) metric at high temperature \cite{Herzog:2006ra,Rinaldi:2021xat}. In here we shall use a unique BH metric for all $T$. The phases will be distinguished by boundary conditions, with the restriction that we shall recover the traditional HW results at $T=0$. A model with a BH metric for low $T$ was used to study the heavy quark potential at finite temperature some time ago \cite{Boschi-Filho:2006hfm}. 

Let us remind the reader that, at $T=0$, glueballs can be described by the gravitons propagating in a $AdS_5$ space governed by the $AdS_5$ metric:

\begin{align}
      ds^2 = \dfrac{L^2}{z^2} \big(dx_\mu dx^\mu-dz^2  \big)
\label{ads5metric}
\end{align}
where $0 \leq z \leq z_0$ {being} $z_0$ the confinement size. The physical particle modes are described by boundary conditions (BC) at  $z_0$.
Here we propose that for $T > 0$  the $AdS$ metric is defined by a BH 

\begin{align}
ds^2 = \frac{L^2}{z^2} \big(f(z) dt^2 - d \vec{x}^2 - f^{-1} (z) dz^2 \big),
\label{bhmetric}
\end{align}
where $f(z)= 1-{z^4}/{z_h^4}$. Note that $z_h$ determines
the Hawking's temperature of the black hole  $T_h = 1/(\pi z_h)$.
For $T=0 \rightarrow z_h=\infty$ with BC at $z_0$ we recover the$AdS_5$ metric, Eq. (\ref{ads5metric}), from the BH metric.

For $T > 0$  the graviton propagates in a BH space, and therefore
the equations of motion for the tensor component in this BH background are~\cite{Constable:1999gb,Brower:2000rp}:

\begin{align}
\frac{d^2\phi(z)}{dz^2} + \left(\frac{2}{z} - \frac{5 z_h^4-z^4}{z
 (z_h^4-z^4)}\right) \frac{d \phi(z)}{dz} + \frac{M^2  z_h^4}{z_h^4-z^4} \phi(z)
= 0,
\label{tensor}
\end{align}
and for the scalar gravitons~\cite{Constable:1999gb,Brower:2000rp}:

\begin{align}
\frac{d^2\phi(z)}{dz^2} + \left(\frac{2}{z} - \frac{5 z_h^4-z^4}{z (z_h^4-z^4)}\right) \frac{d \phi(z)}{dz} + \left(\frac{M^2  z_h^4}{z_h^4-z^4}+ \frac{ 256  z^6 z_h^4}{(z_h^4-z^4)(6 z_h^4-2z^4)^2}\right) \phi(z) = 0.
\label{scalar}
\end{align}

For the sake of simplicity,
 a constant $\lambda$ is introduced in front of the latter Eq. (\ref{scalar})
 so that:
 $\lambda = 1$ corresponds to the scalar graviton and $\lambda=0$ is the equation of the
tensor graviton:

\begin{align}
\frac{d^2\phi(z)}{dz^2} + \left(\frac{2}{z} - \frac{5 z_h^4-z^4}{z
(z_h^4-z^4)}\right) \frac{d \phi(z)}{dz} + \left(\frac{M^2  z_h^4}{z_h^4-z^4}+
\lambda\frac{ 256  z^6 z_h^4}{(z_h^4-z^4)(6 z_h^4-2z^4)^2}\right) \phi(z) = 0.
\label{scalartensor}
\end{align}
Since $z_h$ is related to $T$, the solutions to these equations will provide the dependence of the spectrum with respect to the temperature. In order to solve the above equations to look for stable mode solutions boundary conditions must be imposed. We define two sectors in $z_h$: $i)$ a low temperature sector where $z_h > z_0$, which we will characterize by Dirichlet or Neumann boundary conditions at $z_0$ and $ii)$ a high temperature sector $z_h < z_0$, which satisfies the appropriate boundary conditions at the BH horizon $z_h$. As in Ref. \cite{Rinaldi:2021xat} we would like to find the phase transition region by studying the behavior of the particle modes as a function of temperature. The idea behind our analysis is to control to what extent deconfinement phenomena appear in our calculation as a steep growth in the bound state masses associated with a decrease in binding energy due  to the flattening of the binding potential and therefore leads to the liberation of the valence gluons \cite{Rinaldi:2021dxh}.

\subsection{A two phase model}

Let us assume that we have two phases being $z_0$ the boundary in the fifth dimension $z$. In order to see the behavior of the physical observables at the boundary  we follow the development of Ref. \cite{Herzog:2006ra}. The free energy for a BH metric with the an appropriate ultraviolet cut-off $\varepsilon$ is proportional to

\begin{equation}
V(\epsilon)  \propto -4 \int_0^{\pi z_h} dt  \int_\epsilon^{min(z_0,z_h)} \frac{dz}{z^5} = \pi z_h \left(\frac{1}{(min(z_0,z_h))^4}- \frac{1}{\epsilon^4}\right).
\end{equation}

Let $V_1$ be the free energy for the BH-BC in the confined phase and $V_2$ that for the free BH.  The solutions of the mode equations  for $V_1$  must always satisfy $z_0 < z_h$, while those in $V_2$ have  no restriction. Therefore,  the free energy difference becomes proportional to

\begin{equation}
   \lim_{\epsilon \rightarrow 0} \Delta V (\epsilon) = V_2(\epsilon)-V_1(\epsilon) \propto
\begin{cases}
    0 \quad & \text{if } z_0 < z_h,\\
    \pi z_h(\frac{1}{z_h^4} - \frac{1}{z_0^4})  & \text{if } z_0  > z_h.
\end{cases}
\end{equation}
For  $z_0 > z_h$ only the BH modes are stable, while for $z_0>z_h$ the modes of both solutions are be stable. If we choose for physical reasons a model BH-BC and BH, a first order  phase transition will appear between the two phases with a transition temperature $T=\frac{1}{\pi z_0}$. If we choose BH overall without boundary conditions we will have not phase transition but a crossover. However, choosing overall the BH solution without boundary conditions will lead, as we will show, to zero mass states at $T=0$ and not to the thermal $AdS_5$ spectrum. Therefore,  if we want to recover the correct spectrum at  $T=0$ we are required to have at least a two phase model with a first order phase transition in the dual scheme..

\section{The temperature dependent glueball spectrum with Dirichlet boundary conditions}

We next discuss in detail the procedure to get the temperature dependent scalar glueball spectrum with Dirichlet boundary conditions.

\subsection{The low temperature phase: BH-Dirichlet}
\begin{figure}[htb]
\begin{center}
\includegraphics[scale= 0.9]{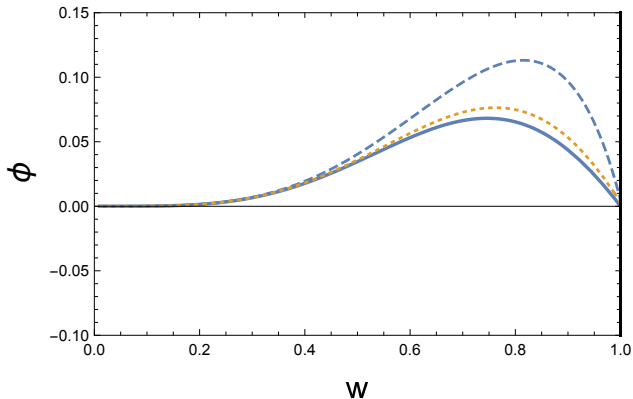} 
\end{center}
\caption{
The scalar glueball mode function
as a function of $w=z/z_h$ for $z_0<z_h$ and Dirichlet boundary conditions. We considered as an example $z_0=1$ and plot the result corresponding to the $AdS_5$ space ($T=0$) (solid), which is related  to large $z_h$, the solution for  $z_h= 1.5$ (dotted) and that for $z_h= 1.1$ (dashed).}
\label{BHDf}
\end{figure}

Let us study the solutions of the equation of motion in the BH background, Eq. (\ref{scalartensor}),  for $z_h >z_0$. 
In particular, 
 the behavior of the equation at the origin $z=0$ can be per properly studied with a change of variable $w=z/z_h$:

\begin{figure}[htb]
\begin{center}
\includegraphics[scale= 0.9]{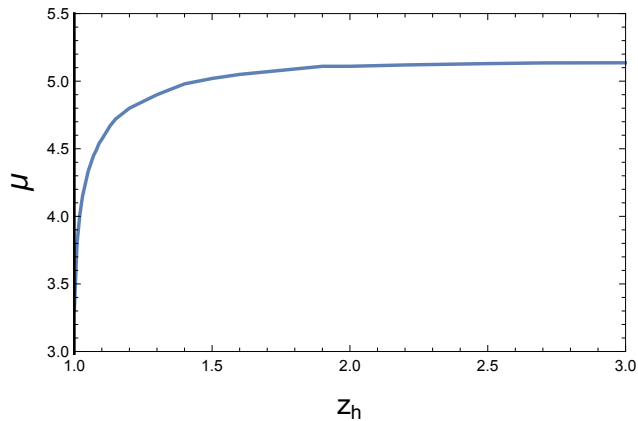} 
\end{center}
\caption{
The scalar glueball modes 
as a function of $z_h$ for $z_0<z_h$ with Dirichlet boundary conditions. The asymptotic value corresponds to the AdS free metric.}
\label{BHDm}
\end{figure}

\begin{align}
\frac{d^2\phi(w)}{dw^2} + \left(\frac{2}{w} - \frac{5  -w^4}{w (1 - w^4)}\right) \frac{d \phi(w)}{dw} + \left(\frac{\tilde{\mu}^2  }{1 - w^4}- \lambda \frac{ 256 w^6}{(1- w^4)(6 -2 w^4)^2}\right) \phi(w) = 0,
\label{bhzhmodes}
\end{align}
where the quantity $\tilde{\mu }= M z_h$. In order to find the modes of
this equation, one needs to integrate from
$w=0$ towards $w_0 \sim z_0/z_h$. The the mode at $w=0$ behaves as

\begin{align}
\phi(w) \sim A w^4 + B
\end{align}
where $A,B$ are integration constants. We take $A=1$ and $B=0$, since different values will change the shape of the mode function but not the mode value, which is our interest at this time.
The behavior of the modes as a function of $z_h$ are shown in fig. \ref{BHDm}. The dropping of the glueball mass close to the phase transition is a consequence of the AdS-BH dynamics of our model and this behavior might be different {within other approaches}  \cite{Vento:2006wh}.
Here we display $\mu$, i.e. the adimensional mass $M/L$ from which the physical one can be obtained by multiplying it for  $L^{-1}=250$ MeV, for Dirichlet boundary conditions \cite{Rinaldi:2021xat}.  As one can see, the mass of the glueballs decreases as the temperature increases.
 For example $\mu= 5.135, 5.00, 4.55$  for $z_h =\infty, 1.5, 1.1$. The asymptotic value corresponds to the asymptotic free metric.
In fig. \ref{BHDf} we show the mode function for different values of $z_h$. The results confirm that as $z_h$ increases, the AdS-BH metric approaches the AdS free one, as expected. {Here, from now on, the numerical evaluations will be performed by fixing $z_0=1$ without losing generality. In fact, our main interest is the functional form of the mode functions and not their sizes.}

\subsection{The high temperature phase}
The next step is to study the solution to Eq. 
(\ref{scalartensor}) for $z_h < z_0$. Boundary conditions must be imposed at the origin and at the horizon. 
One has to match the solutions outward from the origin with the inwards from the horizon to get the mode energies and the mode functions.
To study the behavior of the solution close
to the horizon the following  change of variable is convenient: $ v =1 - w^4$.   Equation (\ref{bhzhmodes}) now becomes:

\begin{align}
\frac{d^2\phi(v)}{dv^2} + \frac{1}{v} \frac{d \phi(v)}{dv} +
\frac{1}{16 v (1-v)^{\sfrac{3}{2}}}\left(\tilde\mu^2  + \lambda \frac{64 (1-v)}{(2 -
v)^2}\right) \phi(v) =0.
\label{bhhorizon}
\end{align}
where $v \rightarrow 0$ at the horizon, i.e. $ w \rightarrow 1$. The regular
solution at $v=0$  has the form of $\phi(v) = \sum_0^\infty a_n v^n$.
Substituting this ansatz into the equation and keeping only terms up to order 3,
one obtains  recurrence relations for $a_i$, with $i \geq 1$, the latter
functions of
the independent
$a_0$
coefficient. For the three first modes  one has:

\begin{align}
a_1 =&- \frac{(16 \lambda +\tilde{\mu}^2 )a_0 }{16}  \nonumber \\
a_2= &\frac{ (16 \lambda - 3 \tilde{\mu}^2 ) a_0 -(32 \lambda + 2 \tilde{\mu}^2 )
 a_1}{128 } \nonumber \\
a_3 =& -\frac{ (80 \lambda + 15 \tilde{\mu}^2 ) a_0 + (64 \lambda - 12 \tilde{\mu}^2 )  a_1
- (128  \lambda + 8 \tilde{\mu}^2 )  a_2}{1152 }.  \nonumber \\
\end{align}

\begin{figure}[htb]
\begin{center}
\includegraphics[scale= 0.9]{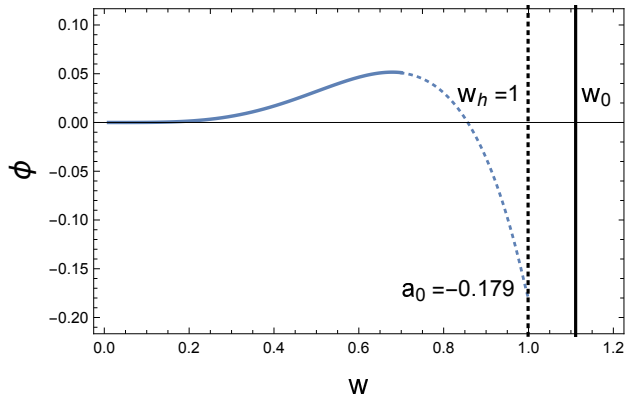} 
\end{center}
\caption{BH mode function for $\tilde{\mu}=5.53$, $z_h=0.9$ and $a_0=-0.179$. The solid curve is the outward solution and the dotted curve is the inward solution.}
\label{BH}
\end{figure}

The approximate solution with these four first terms and its derivative
are used as initial conditions for the numerical program at $v$ close to zero.
In fig. \ref{BH} we show a mode function for mode energy $\tilde{\mu}=5.53, z_h=0.9$ and $a_0 =-0.179$. 
The solid line represents the solution from the origin and the dotted line the inward solution from the horizon $w_h=w(z_h)=1$. $w_0=1/z_h$
represents the confinement radius. The BH solution is now fully stable and contained within the confinement region.

It has been shown \cite{Rinaldi:2021xat} that the mode energy of the BH solution goes like 

\begin{equation}
\mu = \frac{M}{L}=\frac{5.53}{z_h}
\end{equation}
for any $z_h$. 

Let us stress that once the solution is found, as a function of $z_h$, the explicit dependence  on $z_0$ is lost. 
In this case both mode functions and masses do not explicitly depend on $z_0$ and, thus,
the solutions are valid, in principle, for any value of $z_h$. The confinement radius does not impose any restrictions on the BH solutions. The present findings  can be extended for any $z_h$ leading to a crossover at $z_0$. However, the mode energies of these solutions tend to zero asymptotically which does not correspond to phenomenology, see Fig. 
\ref{PhT}. On the contrary the Dirichlet  solutions with BH background, discussed in the previous section, do reproduce the spectrum at zero temperature, within the limitations of the model \cite{Rinaldi:2021xat}.

\subsection{The glueball spectrum as a function of temperature}

\begin{figure}[htb]
\begin{center}
\includegraphics[scale= 0.9]{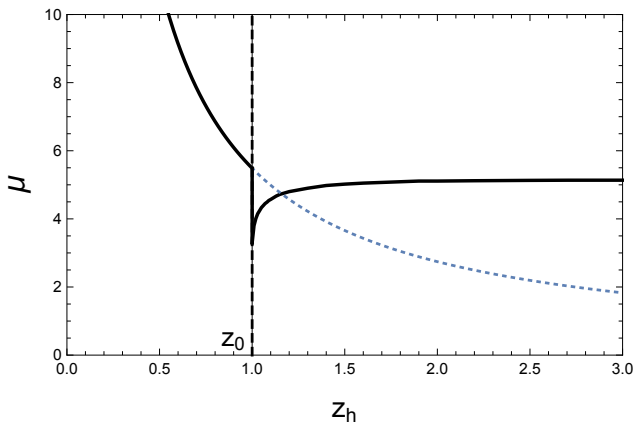} 
\end{center}
\caption{The energy mode associated with the ground 
state scalar glueball as a function of $z_h$, in the BH-Dirichlet and BH scenarios.
The dotted curve represents  a  hypothetical   BH crossover scenario to the low $T$ region. }
\label{PhT}
\end{figure}

In Fig. \ref{PhT} we show the mode energies for the scalar glueball groundstate in the two scenarios BH-Dirichlet  and BH  as a function of $z_h$; recall that  $T_h\sim \frac{1}{\pi z_h}$. The BH mode function is continuous at the confinement boundary, while the BH-Dirichlet mode function is only defined for $z_h> z_0$. We se that the former tends to $0$ for $T=0$, while the latter tends to the $AdS_5$ thermal value at $T=0$. Therefore a two phase model is motivated by the data, with the BH geometry for $z_h<z_0$ and the BH-Dirichlet geometry for $z_h>z_0$. In these model the two energy mode functions have a jump at the confinement boundary $z_0$, thus manifesting the fisrt order phase transition described above. This value of $z_0$ corresponds in this model to $T =\frac{1}{\pi z_0} \sim 80$ MeV since $L^{-1}=250$ MeV for Dirichlet boundary conditions \cite{Rinaldi:2021xat}. It is well known that the HW models lead to low transition temperatures. 

The BH phase describes the deconfined phase in our model.  In the confined region the mass of the modes is mostly determined by the boundary conditions at $z_0$, while in the deconfined region it is defined by the boundary conditions at the BH horizon $z_h$. Thus, in our model based on the HW approach the phase transition is determined by boundary conditions and not by the dynamics, i.e. the phases change due to the boundary conditions and not due to a change of the background metric.

\begin{figure}[htb]
\begin{center}
\includegraphics[scale= 0.9]{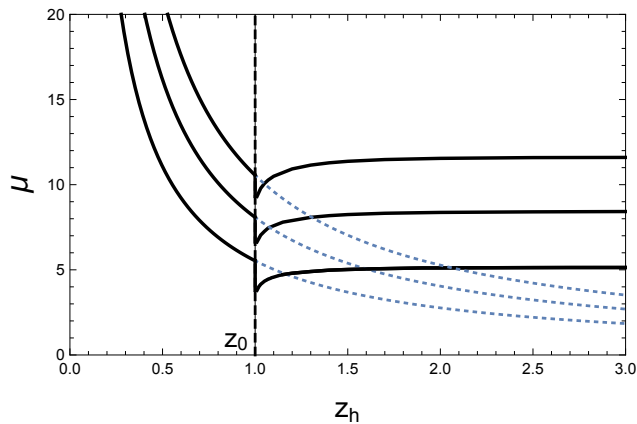} 
\end{center}
\caption{The energy modes associated with the ground state and two excited states as a function of $z_h$ for the scalar glueball in the  BH-Dirichlet and BH scenarios (solid) showing the phase transition. { The dotted curve represents  a  hypothetical   BH crossover scenario to the low $T$ region.} }
\label{PhTDexcited}
\end{figure}

In Fig. \ref{PhTDexcited} we show the effect of temperature on the excited scalar glueball states. We note that the same phenomenon, observed for the ground state, takes place. In the two phase model a jump in the modes occurs at the transition temperature. Also in these cases one has a non physical crossover scenario.

\begin{figure}[htb]
\begin{center}
\includegraphics[scale= 0.9]{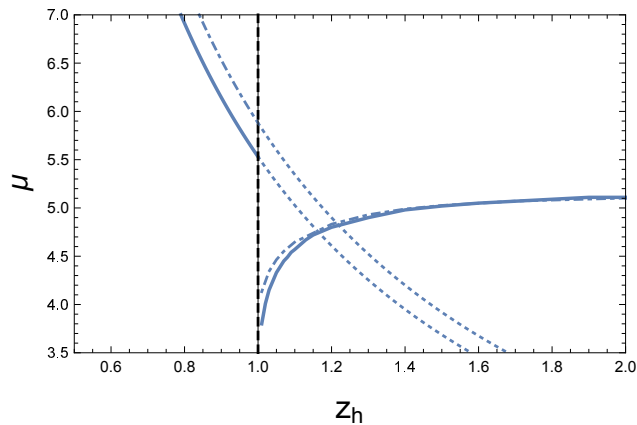} 
\end{center}
\caption{We show the energy modes associated with the ground state scalar and tensor glueballs as a function of $z_h$ in the  BH-Dirichlet  and in the BH scenarios (solid) and signal the {hypothetical} crossover phenomenon by a dotted curve. }
\label{PhTDtensor}
\end{figure}

In order to conclude the analysis for the Dirichlet boundary conditions we study the tensor glueballs.
Let us remark that the HW model at zero temperature predicts 
that scalar and tensor glueballs are degenerate  \cite{Rinaldi:2017wdn}. However, since in the BH metric the tensor equation differs from the scalar one, recall Eq. (\ref{scalartensor}),
 once we heat the system their masses will be different, a phenomenon that was also addressed in our previous study of the Herzog type phase transition \cite{Rinaldi:2021dxh}. In Fig. \ref{PhTDtensor} we show the phase transition for the scalar and tensor glueballs. We stress  that the modes are only slightly different at low temperatures because the
{term proportional to}
$\lambda$ in Eq. (\ref{scalartensor}) is small there,  but it becomes quite distinct at high temperatures. 

\begin{figure}[htb]
\begin{center}
\includegraphics[scale= 0.9]{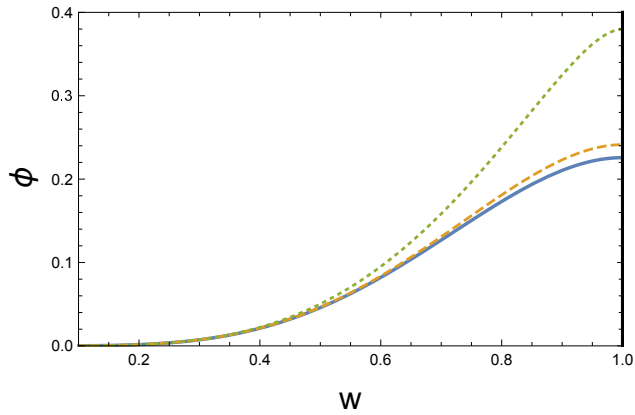} 
\end{center}
\caption{The functional modes associated with the ground state scalar  as a function of $w$ for three values of $z_h =2.0$ (dotted) , $1.5$ (dashed), $1.1$ (solid). }
\label{Nmodes}
\end{figure}

\section{The temperature dependent glueball spectrum  with Neumann boundary conditions}
\begin{figure}[htb]
\begin{center}
\includegraphics[scale= 0.9]{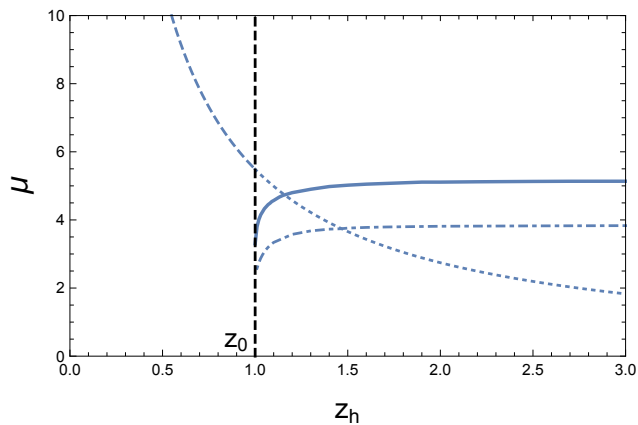} 
\end{center}
\caption{The energy modes associated with the scalar glueball as a function of $z_h$, for the Dirichlet (dotdashed) and Neumann (solid) at low temperature  and in the BH scenario (dashed), which is equal in both scenarios. We also signal the { hypothetical} crossover phenomenon by a dotted curve. }
\label{PhTDN}
\end{figure}

In this section we describe the dynamics by 
imposing  Neumann boundary conditions  at the confinement boundary $z_0$ to the BH equations of motion, Eq.(\ref{scalartensor}). The resulting mode functions are shown in Fig. \ref{Nmodes}. The latter,  for $z_h< z_0$, 
coincide with those 
already  shown in
Fig. \ref{BH}. The relative mode energies for the hight T region correspond to those shown in
 Fig. \ref{PhT} for $z_h<z_0$. In Fig. \ref{PhTDN} we plot the variation of the mode energy as a function of $z_h$ comparing with the Dirichlet case . The mode energies vary slowly decreasing as $z_h$ decreases up to the confinement boundary $z_0$ and  grow dramatically as we go below the confinement boundary separating  the two scenarios, completely analogous to the Dirichlet case as shown in Fig. \ref{PhTDN}. The solid curve represents  transition of the mode  function in the model. The crossing occurs at $z_0$ which corresponds to a temperature $T \sim 92$ MeV, since $L^{-1} = 290$ MeV for the Neumann boundary conditions \cite{Rinaldi:2021xat}. We note that this temperature is again low.
 One could proceed by studying excitations and the tensor components but the behavior of the mode function would be similar as before, only the magnitudes would change.

\section{Analysis of the transition}
{In this section, we}
 analyze the phase transition obtained in our AdS-BH model from the perspective of {two different} constituent gluon models.

\subsection{Constituent Gluons}

We first {consider} a constituent model of two heavy gluons of mass $m$ coupled by a confining potential of the form

\begin{equation}
V(r) = 2m(1-e^{-\sfrac{r}{r_s}}).
\label{potential}
\end{equation}
which we shall call Cornwall-Soni model~\cite{Cornwall:1982zn}.

By introducing this potential we are assuming that the glueball is confined bellow $2m$. In this case the reduced Sch\"odinger equation can be written as

\begin{equation}
\frac{d^2 \chi}{d r^2} = m(V(r)-E)) \chi,
\label{Sch}
\end{equation}
where the wave function $\Psi$ is related to the reduced wave function $\chi$ by $\Psi =\frac{\chi}{r}$.
This equation will provide the glueball masses for each temperature by relating the potential to $z_h$ equating the glueball masses obtained from the solution of the Schr\"odinger equation to those of the $AdS_5$ model. We do this  analysis for Dirichlet boundary conditions only.  To be more specific, we take for $z_h$ the corresponding mode value and transform it into a gluon mass $M_G$ dividing by $L^{-1} = 250$ MeV. We then solve Eq.(\ref{Sch}) for different values of $r_s$ until $E=M_G$. In this way we find a functional form of the potential parameter $r_s$ in terms of $z_h$. We assume that the gluon mass $m$ does not change with temperature and take the value $m=1$ GeV used in ref.~\cite{Cornwall:1982zn}.  In Fig. \ref{strength} we show the functional relation of $r_s$ on $z_h$ both for the Hawking-Page transition model (HP)~\cite{Rinaldi:2021dxh,Herzog:2006ra} (dotted) and for the present model (BHD) (solid). The HP model has a phase transition at $z_{HP}= \frac{z_0}{2^{\sfrac{1}{4}}}$, while the BHD model the transition at $z_h=z_0$, as proven. However, the two models coincide beyond the Herzog temperature \cite{Herzog:2006ra}  (dotdashed).

\begin{figure}[htb]
\begin{center}
\includegraphics[scale= 0.9]{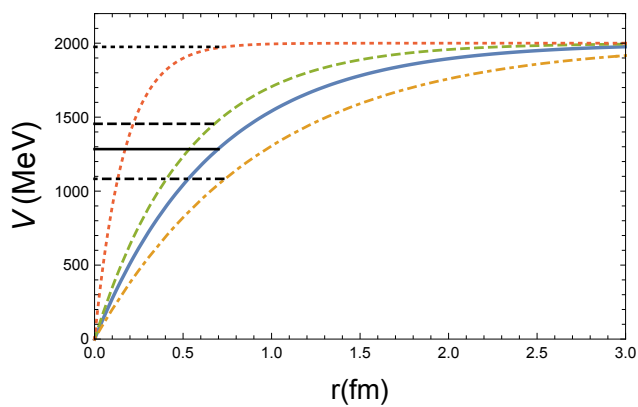} 
\end{center}
\caption{We show the confinement potential for four  temperatures. For zero temperature (solid), slightly below the transition point (dotdashed), slightly above the phase transition (dashed) and for very large temperature with small binding energy (dotted). The lines represent the glueball masses for each potential with the same symbols.}
\label{ConfPot}
\end{figure}

In Fig. \ref{ConfPot} we describe  how the confinement potential of the Cornwall-Soni model is altered as we increase the temperature. To this aim, we plot the potential for four values of $z_h$: AdS Thermal (solid), $z_h= 1.05 z_0 $  slightly to the right of the transition temperature (dotdashed), $z_h= 0.95$ slightly to the left of the transition point (dashed), and finally for very high temperature (dotted). As we approach the transition temperature from below the binding becomes stronger and the glueball mass decreases. At the transition temperature there is a jump and the mass of the glueball increases above the thermal value and from then on keeps on increasing until for very high temperature the binding becomes vanishingly small and the mass of the glueball become $2m$, that of two free gluons. The potential becomes wider before the phase transition increasing the binding energy, and narrower after the phase transition decreasing the binding energy, until the mass of the glueball becomes that of two free gluons.

\begin{figure}[htb]
\begin{center}
\includegraphics[scale= 0.9]{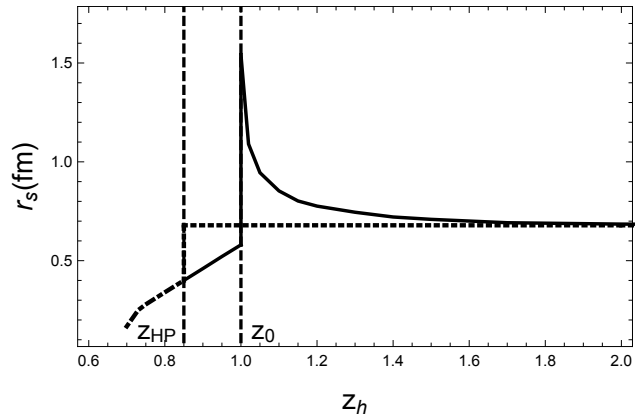} 
\end{center}
\caption{The dependence of the potential strength parameter $r_s$ as a function of $z_h$ for the HP (dotted) before the phase transition for the CM  (dashed) and for both models when they are equal (dotdashed).}
\label{strength}
\end{figure}

\begin{figure}[htb]
\begin{center}
\includegraphics[scale= 0.9]{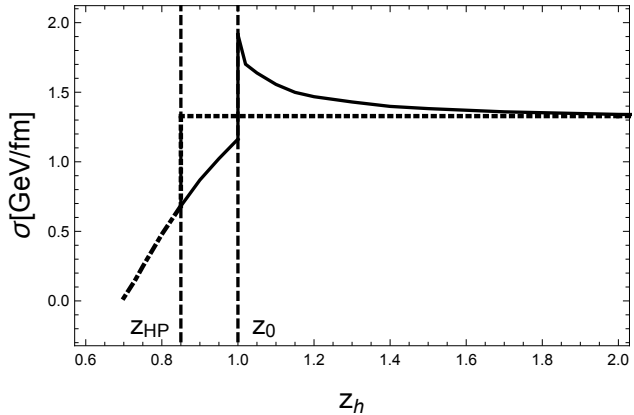} 
\end{center}
\caption{The behavior of the string tension for the HP (dotted) and BHD (dashed) models when they differ and (dotdashed) when they coincide. }
\label{sigma}
\end{figure}

Let us try to understand this deconfinement behavior in terms of a string tension, i.e.,  two gluons kept together by a flux tube. We define the string tension in this potential model as

\begin{equation}
\sigma = \frac{|E_B|}{R},
\label{stringtension}
\end{equation}
 {where $E_B$ is the binding energy and $R=\int d^3r \Psi^* r^2 \Psi$ is the mean square radius of the glueball state $\Psi$ for each mass}.  In Fig. \ref{sigma} we plot the string tension as a function of $z_h$. In the BHD model the string tension decreases after the transition temperature and coincides with that of the HP model after the Herzog temperature. The string tension is discontinuous at the transition temperature, decreasing thereafter and vanishing after total deconfinement.

It is important to point out that in the potential model the Schr\"odinger equation ceases to have solutions for values of the glueball mass above $2m$, as expected in this simple non relativistic approach. Therefore,  in defining the shape of the potential, Eq.(\ref{potential}), we have introduced a cutoff in the AdS high temperature spectrum to eliminate the glueballs with masses above $2m$, which appear in our formalism. In fact, the holographic model has glueball solutions for temperatures above the transition temperature both  the HP and the BHD in line with the arguments of Shuryak and Zahed \cite{Shuryak:2003ty}.

{
\subsection{Cornell Potential}

Let us analyze {the} results with another potential scheme. We study a  Cornell  potential

\begin{equation}
V(r) = a r- 3 \frac{\alpha_s}{r}
\label{Cornell}
\end{equation}

that has been used within a semirelativistic 
approach \cite{Brau:2004xw,Buisseret:2006da,Mathieu:2008bf}.
 The treatment of the semirelativistic hamiltonian requires complicated mathematical tools \cite{Buisseret:2004ag}, 
which would obscure our analysis. {Hence}, we have adopted  a
 non relativistic scheme with an effective gluon mass $\mu$ 
which leads to equation (\ref{Sch}) with the Cornell potential Eq.
(\ref{Cornell}). The parameters of the potential in the cited 
calculations have the following interpretation: $a =C\sigma$ 
where $C=9/4$ is a color factor and $\sigma=0.185$ GeV$^2$ is 
the string tension of the mesonic flux tube;  $\alpha_s=0.2$ is
 the asymptotic QCD coupling constant. These authors find an
 effective mass of about $600$ MeV. From their experience we  
take for  our non relativistic model $\mu=600$ MeV , $\alpha_s=0.2$,
 $C=9/4$, but we need a value of $\sigma \sim 0.2$ Gev$^2$, which is
 close to theirs, to fit our AdS-HW  $T=0$ results. In order to 
implement the temperature dependence we will proceed as before we 
vary $a$, the gluon flux tube string tension, in order to fit our AdS 
$T\ne0$ glueball mass values. We  keep the effective mass and the 
asymptotic coupling constant fixed. The result of our calculation can
 be looked up in Fig. \ref{strength}, where we also compare with the 
HP transition.

\begin{figure}[htb]
\begin{center}
\includegraphics[scale= 0.7                                                                                                                                                                                                                                                                                                                                                                                                                                                                                                                                                                                                                                                                                                                                                                                                                                                                                                                                                                                                                                                                                                                                                                                                                                                                                                                                                                                                                                                                                                                                                                                                                                                                                                                                                                                                                                                                                                                                                                                                                                                                                                                                                                                                                                                                                                                                                                                                                                                                                                                                                                                                                                                                                                                                                                                                                                                                                                                                                                                                                                                                                                                                                                                                                                                                                                                                                                                                                                                                                                                                                                                                                                                                                                                                                                                                                                                                                                                                                                                                                                                                                                                                                                                                                                                                                                                                                                                                                                                                                                                                                                                                                                                                                                                                                                                                                                                                                                                                                                                                                                                                                                                                                                                                                                                                                                                                                                                                                                                                                                                                                                                                                                                                                                                                                                                                                                                                                                                                                                                                                                                                                                                                                                                                                                                                                                                                                                                                                                                                                                                                                                                                                                                                                                                                                                                                                                                                                                                                                                                                                                                                                                                                                                                                                                                                                                                                                                                                                                                                                                                                                                                                                                                                                                                                                                                                                                                                                                                                                                                                                                                                                                                                                                                                                                                                                                                                                                                                                                                                                                                                                                                                                                                                                                                                                                                                                                                                                                                                                                                                                                                                                                                                                                                                                                                                                                                                                                                                                                                                                                                                                                                                                                                                                                                                                                                                                                                                                                                                                                                                                                                                                                                                                                                                                                                                                                                                                                                                                                                                                                                                                                                                                                                                                                                                                                                                                                                                                                                                                                                                                                                                                                                                                                                                                                                                                                                                                                                                                                                                                                                                                                                                                                                                                                                                                                                                                                                                                                                                                                                                                                                                                                                                                                                                                                                                                                                                                                                                                                                                                                                                                                                                                                                                                                                                                                                                                                                                                                                                                                                                                                                                                                                                                                                                                                                                                                                                                                                                                                                                                                                                                                                                                                                                                                                                                                                                                                                                                                                                                                                                                                                                                                                                                                                                                                                                                                                                                                                                                                                                                                                                                                                                                                                                                                                                                                                                                                                                                                                                                                                                                                                                                                                                                                                                                                                                                                                                                                                                                                                                                                                                                                                                                                                                                                                                                                                                                                                                                                                                                                                                                                                                                                                                                                                                                                                                                                                                                                                                                                                                                                                                                                                                                                                                                                                                                                                                                                                                                                                                                                                                                                                                                                                                                                                                                                                                                                                                                                                                                                                                                                                                                                                                                                                                                                                                                                                                                                                                                                                                                                                                                                                                                                                                                                                                                                                                                                                                                                                                                                                                                                                                                                                                                                                                                                                                                                                                                                                                                                                                                                                                                                                                                                                                                                                                                                                                                                                                                                                                                                                                                                                                                                                                                                                                                                                                                                                                                                                                                                                                                                                                                                                                                                                                                                                                                                                                                                                                                                                                                                                                                                                                                                                                                                                                                                                                                                                                                                                                                                                                                                                                                                                                                                                                                                                                                                                                                                                                                                                                                                                                                                                                                                                                                                                                                                                                                                                                                                                                                                                                                                                                                                                                                                                                                                                                                                                                                                                                                                                                                                                                                                                                                                                                                                                                                                                                                                                                                                                                                                                                                                                                                                                                                                                                                                                                                                                                                                                                                                                                                                                                                                                                                                                                                                                                                                                                                                                                                                                                                                                                                                                                                                                                                                                                                                                                                                                                                                                                                                                                                                                                                                                                                                                                                                                                                                                                                                                                                                                                                                                                                                                                                                                                                                                                                                                                                                                                                                                                                                                                                                                                                                                                                                                                                                                                                                                                                                                                                                                                                                                                                                                                                                                                                                                                                                                                                                                                                                                                                                                                                                                                                                                                                                                                                                                                                                                                                                                                                                                                                                                                                                                                                                                                                                                                                                                                                                                                                                                                                                                                                                                                                                                                                                                                                                                                                                                                                                                                                                                                                                                                                                                                                                                                                                                                                                                                                                                                                                                                                                                                                                                                                                                                                                                                                                                                                                                                                                                                                                                                                                                                                                                                                                                                                                                                                                                                                                                                                                                                                                                                                                                                                                                                                                                                                                                                                                                                                                                                                                                                                                                                                                                                                                                                                                                                                                                                                                                                                                                                                                                                                                                                                                                                                                                                                                                                                                                                                                                                                                                                                                                                                                                                                                                                                                                                                                                                                                                                                                                                                                                                                                                                                                                                                                                                                                                                                                                                                                                                                                                                                                                                                                                                                                                                                                                                                                                                                                                                                                                                                                                                                                                                                                                                                                                                                                                                                                                                                                                                                                                                                                                                                                                                                                                                                                                                                                                                                                                                                                                                                                                                                                                                                                                                                                                                                                                                                                                                                                                                                                                                                                                                                                                                                                                                                                                                                                                                                                                                                                                                                                                                                                                                                                                                                                                                                                                                                                                                                                                                                                                                                                                                                                                                                                                                                                                                                                                                                                                                                                                                                                                                                                                                                                                                                                                                                                                                                                                                                                                                                                                                                                                                                                                                                                                                                                                                                                                                                                                                                                                                                                                                                                                                                                                                                                                                                                                                                                                                                                                                                                                                                                                                                                                                                                                                                                                                                                                                                                                                                                                                                                                                                                                                                                                                                                                                                                                                                                                                                                                                                                                                                                                                                                                                                                                                                                                                                                                                                                                                                                                                                                                                                                                                                                                                                                                                                                                                                                                                                                                                                                                                                                                                                                                                                                                                                                                                                                                                                                                                                                                                                                                                                                                                                                                                                                                                                                                                                                                                                                                                                                                                                                                                                                                                                                                                                                                                                                                                                                                                                                                                                                                                                                                                                                                                                                                                                                                                                                                                                                                                                                                                                                                                                                                                                                                                                                                                                                                                                                                                                                                                                                                                                                                                                                                                                                                  ]{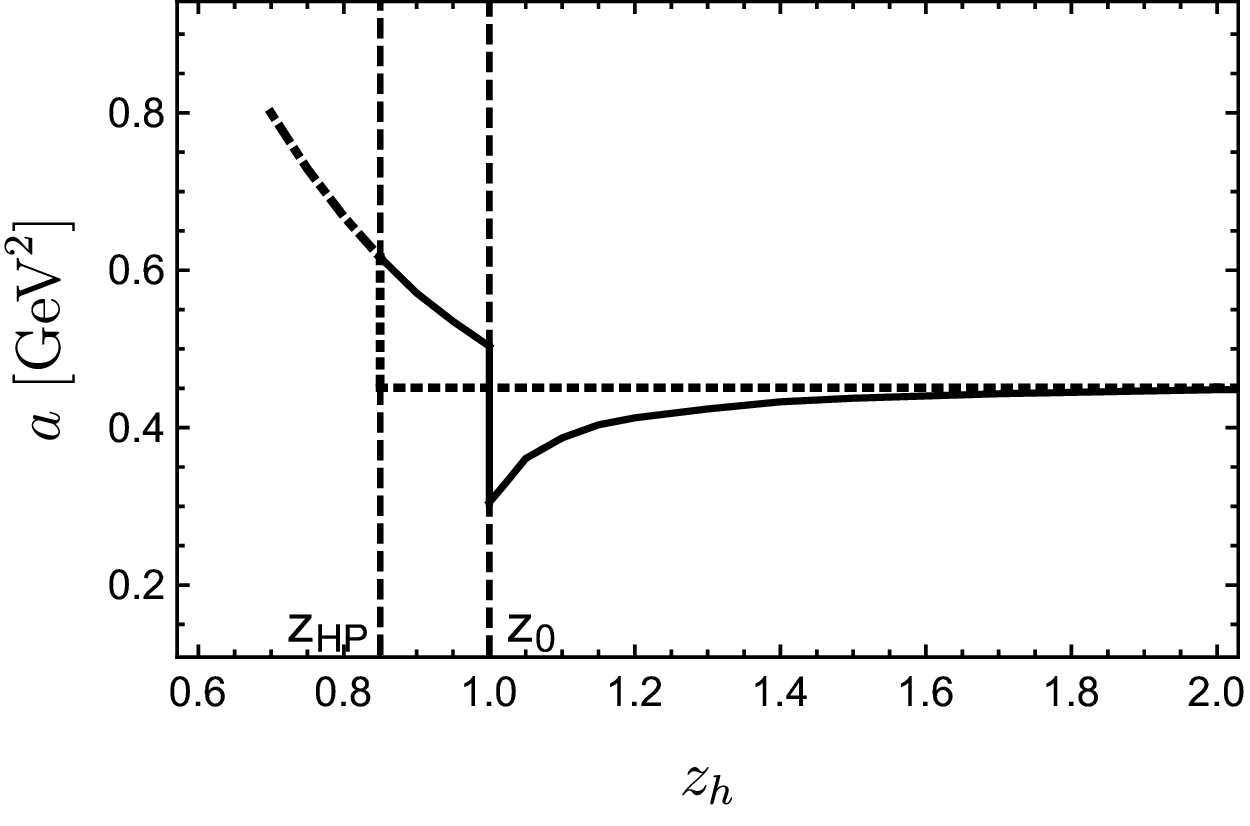} 
\end{center}
\caption{The dependence of the potential fluxtube parameter $a$ as a function of $z_h$ for the HP (dotted) and for the Cornell  (dashed) before the phase transition, and for both models when they are equal (dotdashed).}
\label{strength}
\end{figure}

\begin{figure}[htb]
\begin{center}
\includegraphics[scale= 0.9]{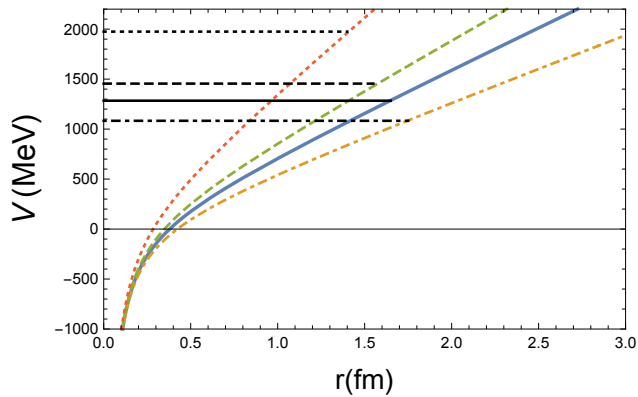} 
\end{center}
\caption{We show the confinement potential for four  temperatures. For zero temperature (solid), slightly below the transition point (dotdashed), slightly above the phase transition (dashed) and for very large temperature with small binding energy (dotted). The lines represent the glueball masses for each potential with the same symbols.}
\label{ConfPotCornell}
\end{figure}

In Fig. \ref{ConfPotCornell} we describe  how the confinement potential of the Cornell model is altered as we increase the temperature. To this aim, we plot the potential for four values of $z_h$: AdS Thermal (solid), $z_h= 1.05 z_0 $  slightly to the right of the transition temperature (dotdashed), $z_h= 0.95$ slightly to the left of the transition point (dashed), and finally for very high temperature (dotted). The behavior is very similar to that of the Cornwall-Soni potential. As we approach the transition temperature from below the binding becomes stronger and the glueball mass decreases. At the transition temperature there is a jump and the mass of the glueball  increases above the thermal value and from then on keeps on increasing. The potential becomes wider before the phase transition {thus} increasing the binding energy, and narrower after the phase transition {hence} decreasing the binding energy. 

In this case  neither the Cornell potential nor our AdS-BH dynamics have a confinement mechanism at large $T$.  In order to introduce it we  impose a threshold in the potential.
The main difference between the Cornell and Cornwall-Soni potential is that the latter has a natural threshold at $2m$. 

Lattice data show scalar glueballs up to $4000$ MeV \cite{Morningstar:1999rf,Chen:2005mg,Lucini:2004my} . Thus, the interpretation of the effective mass of the gluon by Cornwall-Soni potential seems only valid for the lowest scalar glueball, whose mass is below $2m$. The Cornell potential does not have a natural threshold, therefore we will introduce a threshold by fiat, which is a new ingredient in our description of confinement.

Let us describe deconfinement again in terms of a string tension, Eq. (\ref{stringtension}). In this case we introduce a threshold $E_{TH}$ and define the binding energy

\begin{equation}
E_B= M_G - E_{TH}.
\end{equation}
Again $R$ is the mean square distance between the gluons. 

  In Fig. \ref{thresholds} we plot the string tension as a function of $z_h$ for several thresholds. We do not plot the behavior of the HP model because is similar to that of the previous potential, constant at the $T=0$ value up to $z_H=0.85$ and then equal to that of the BDH model.  The behavior for the BHD model is as follows: the string tension is almost constant in the confined region except close to the phase transition, where it increases slightly. After the phase transition  in the deconfined region it decreases continuously. We notice that the string tension depends crucially on the thresholds increasing when we increase  the thresholds.

\begin{figure}[htb]
\begin{center}
\includegraphics[scale= 0.9]{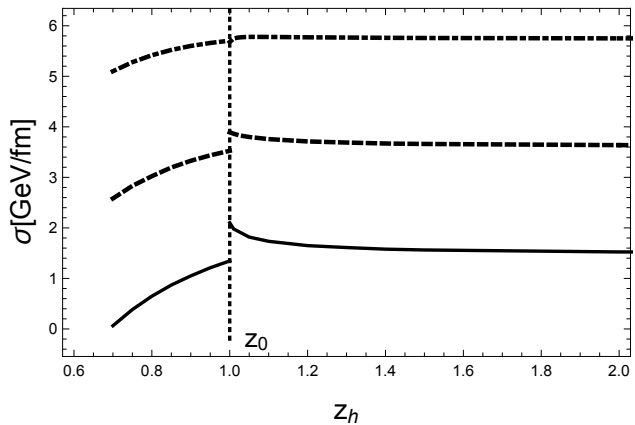} 
\end{center}
\caption{The behavior of the string tension for  the BHD for the following thresholds: $E_{TH}= 2000$ (solid), $E_{TH}= 3000$ (dashed) , $E_{TH}= 4000$ (dotdashed).}
\label{thresholds}
\end{figure}

Let us recapitulate, by concluding that to this order in the $1/N$ expansion, the AdS model leads to an infinite bound potential, without any threshold. We can interpret these results as the creation of colored gluon glueballs at high temperatures without deconfinement~\cite{Shuryak:2003ty}, and therefore, this specific holographic model does not allow to reach the Quark Gluon Plasma. 
}

\section{Conclusions}
\label{conclusions}
Holographic models describe the QCD deconfinement transition as a Hawking-Page  transition from AdS thermal to a BH phase~\cite{Herzog:2006ra}. We recently investigated how this phase transition affects the glueball spectrum in the HW model where the scalar and tensor glueballs are described as gravitons~\cite{Rinaldi:2021dxh}. We found that the glueball mass grows rapidly as we increase the temperature above the phase transition temperature. In our approach the bound state structure does not disappear, and the mass of th glueballs becomes higher than that of a fixed number of free valence gluons. The model thus requires the implementation of a second phase transition  to describe the Quark Gluon Plasma. The resulting scenario is similar to a phase transition a la Shuryak-Zahed where massive, even colored states could appear before total deconfinement~\cite{Shuryak:2003ty}. 
These HP models  lead to a first order phase transition to deconfinement,  with no temperature dependence before the phase transition.

In this work we have followed the same approach as in our previous investigation by studying the behavior of the glueball states as we increase temperature. We use again  an HW type model because it simplifies the study considerably and is very indicative of how holography describes the transition, despite the fact that they are quatitatively not very successful. The main difference with respect to our previous calculation~\cite{Rinaldi:2021dxh} is that we use the BH metric for all values of the temperature. For the BH radius $z_h$ greater than the confinement radius $z_0$, i.e. for small temperatures, we force the solution to be secluded into the physical volume $0<z<z_0$ by imposing on the BH solution Dirichlet or Neumann boundary conditions at $z_0$. In this way we introduce temperature dependence in the confining region. For $z_h<z_0$ we solve the BH solution up to the BH radius $z_h$, however the BH solution exists also beyond $z_0$. Recall that $T \sim \frac{1}{\pi z_h}$.

The solution for the first scenario coincides asymptotically, for $z_h \rightarrow\infty$, with the AdS thermal solution. When the temperature increases the mass of the glueballs decrease very slowly until $z_h=z_0$ where they reach a finite value (see Fig. \ref{PhT}). On the other hand the BH energy mode diverges at the origin $z_h \rightarrow 0$ and decreases towards zero at $z_h\rightarrow \infty$. At $z_0$ there is a phase transition which we have proven is first order.

In the new scheme we see the same phenomenon arising as in our previous calculation, namely an increase in the glueball masses associated with the reduction of the gluon interaction that leads to higher glueball masses. Since our model is equivalent to a rising potential the glueball masses tend to infinity at very high temperature. It is clear that the softening of the potential and the deconfinement of the gluons is not contemplated in the model. The model points, as in our previous calculation, to a complicated bound state scenario after the transition temperature leading to QGP at higher temperatures.  The gluons at some temperature are escaping the well as their glueball masses become larger than the constituent gluon masses. The higher the initial glueball masses  the sooner the gluons will be liberated.

{In order to understand the AdS models in physical terms we have introduced  potential models to determine the mass of a glueball formed by two valence gluons. {The first}
 one  is constructed to produce deconfinement at $2m$ \cite{Cornwall:1982zn}, $m$ being the constituent gluon mass. The other model, inspired by relativistic calculations, is defined by a Cornell type potential. We have forced the AdS glueball masses into these potentials finding in this way the dependence of the potential parameters with temperature. We have described in this way the HP and the BH transitions for the string tension, which vanishes at  deconfinement. We see that glueballs still exist beyond the transition temperatures only ceasing to exist by fiat when they reach the deconfinement threshold. Thus in this way of proceeding we have generated a model with two transitions, one governed by the AdS dynamics and a second by the imposed threshold.}

\section*{Acknowledgements}
 This work was supported in part  by $i)$ Ministerio de Ciencia e Innovaci\'on and Agencia Estatal de Investigaci\'on of Spain MCIN/AEI/10.13039/501100011033, European Regional Development Fund Grant No. PID2019-105439 GB-C21 and by GVA PROMETEO/2021/083; $ii$ the European Union's Horizon 2020 research and innovation programme under
grant agreement STRONG - 2020 - No 824093. VV would like to thank Vincent Mathieu for useful discussions regarding the calculations within the semirelativistic model and to thank the hospitality of the INFN-Perugia during the development of this investigation.

\bibliographystyle{unsrt}

\bibliography{GravitonT.bib}

\end{document}